\documentclass[twocolumn,showpacs,preprintnumbers,amsmath,amssymb,showkeys,nofootinbib]{revtex4}

\usepackage{ifpdf}
\usepackage{graphicx}
\usepackage{dcolumn}
\usepackage{bm}
\usepackage{aas_macros}

\usepackage{mdwlist}
\usepackage{hyperref}   
\newcommand{\rmbf}[1]{{\rm \bf{#1}}}

\begin{document}

\preprint{APS/123-QED}

\title{The Kinetic Sunyaev-Zel'dovich effect of the Milky Way Halo}

\author{Yuval Birnboim}
\affiliation{Harvard Smithsonian CfA. 60 Garden Street, Cambridge, MA,
  02138, USA}
 \email{ybirnboim@cfa.harvard.edu}
\author{Abraham Loeb}%
\affiliation{Harvard Smithsonian CfA. 60 Garden Street, Cambridge, MA,
  02138, USA}

\date{\today}

\begin{abstract}
We calculate the expected imprint of the ionized gas in the Milky-Way
halo on the Cosmic Microwave Background (CMB) through the kinetic
Sunyaev-Zel'dovich (kSZ) effect. Unlike other Galactic foregrounds,
the halo kSZ signature covers the full sky, generates anisotropies on
large angular scales, is not accompanied by spectral distortions, and
could therefore be confused with primordial CMB anisotropies.  We
construct theoretical models for various halo components, including
smooth diffuse gas, filaments of cold inflowing gas and high velocity
clouds.  We find that the kSZ effect for all components is above the
sensitivity of the {\it Planck} satellite, over a range of angular
scales.  However, the typical halo contribution is well below the
cosmic variance noise in the primordial CMB power spectrum. High
velocity clouds could dominate the halo contribution and better
observational data is required to mask them out. We derive expected
kSZ maps based on existing data from tracers of the halo gas
distribution, such as 21cm maps of neutral hydrogen and H$_\alpha$
maps of recombining gas. The cross-correlation of these maps with the
WMAP5 data does not yield any statistically significant signal.
\end{abstract}


\maketitle

\section{Introduction}
\label{sec:introduction}

The scattering of the Cosmic Microwave Background (CMB) on free
electrons with a bulk peculiar velocity relative to the cosmic rest
frame, shifts the observed CMB temperature and induces the so-called
{\it kinetic Sunyaev-Zel'dovich (kSZ) effect}
\citep[kSZ,][]{sz80}. Unlike its thermal counterpart which (for an
isotropic distribution of electron velocities) alters the shape of the
CMB spectrum, the kSZ effect shifts the spectrum while maintaining its
blackbody shape. This makes the kSZ signal on large angular scales a
foreground contamination that is indistinguishable from the CMB
anisotropies induced at the last scattering surface located at the
redshift of cosmological recombination, $z\sim 10^3$.

The removal of the kSZ foreground requires knowledge of the
distribution of free electrons and their velocity pattern.  A local
system of ionized gas that covers the entire sky and is capable of
inducing kSZ anisotropies on large angular scales is the halo of our
own Milky-Way (MW) galaxy.  In this paper, we examine the kSZ signature
imprinted by the ionized halo gas. Our primary goal is to find whether
this unknown kSZ foreground is an important contaminant which limits
our ability to extract the primordial anisotropies on the CMB sky.

The MW galaxy can be modelled as a typical disk galaxy in the
concordance Cold Dark Matter ($\Lambda$CDM) Universe.  Theoretical
studies of galaxy formation \citep{bd03,keres05,db06} predict that in
the most massive halos with $\gtrsim 10^{12}M_\odot$, the infalling
gas is shock-heated to the virial temperature of the halo, where it
resides in hydrostatic equilibrium until cooling allows it to drift
inwards and make stars in a central galaxy.  This configuration is
obtained in hydrodynamic simulations and observed in X-ray
clusters. Lower-mass halos are too cold to emit hard X-rays, and it is
challenging to observe their soft X-ray emission.  The ionized gas in
cluster halos causes a kSZ contamination of the CMB which is smaller
than its thermal counterpart \citep{carlstrom02}.
\citet{dolag05} estimated the SZ effect of local large scale
structures using constrained realization simulations and found the kSZ
to be sub-dominant to the thermal SZ, both peaking at $l\sim 1000$.
  \citet{suto96}
suggested that the thermal effect of the local group might be
detectable in CMB measurements \citep[see, however, Ref. ][which claims
that the assumed local group halo contradicts observations of other
galaxy groups]{pildis96}. Other systems of ionized gas, including the
intergalactic medium during the reionization era, also provides a
negligible contribution on large angular scales \citep[for example,
Ref. ][ finds that the kSZ contribution from reionization peaks on the
small angular scale of a few arcminutes]{mcquinn05}.

Although the existence of gaseous halos around galaxies is a robust
theoretical prediction, its direct observation has so far proven to be
challenging \citep{maloney99}.  \citet{nicastro02} and \citet{rasmussen03} both find
evidence of extended halo around the local group containing as much as
$10^{12}M_\odot$ and $10^9-10^{11}M_\odot$ of gas respectively, by
using OVII and OVIII absorption lines (and assuming a metallicity
$Z/Z_\odot =0.3$--$1$).  Other studies used the emission of OVII
\citep{sanders02,bregman07a}, and the difference between the absorption
line strength of the local group and the MW \citep{bregman07a} to argue
that the gas is more likely to be bound to the MW than the local
group. 
This conclusion is consistent with \citet{sembach03,yao08} which
detect OVI lines but do not detect OVII line indicating that the gas
is bound to the shallower MW potential well.
Hydrodynamic constraints from the interaction between the
Magellanic stream gas and the ambient dilute gas predict a
non-negligible gaseous halo at a distance of $\sim 50{\rm kpc}$ from
the center of the MW \citep{weiner96,putman03}.  The existing results (see
review in Ref. \citep{bregman07b}) are sensitive to uncertain
assumptions about the metalicity and the existence of a
multi-temperature or multi-component gas.  We therefore follow a
theoretical approach in modeling the gas in the MW halo and the local
group.  The latest models \citep{reid09,shattow08} of the MW predict
that it has a total mass of $M_{virial}\approx 1-2 \times
10^{12}M_\odot$, a baryonic mass of $M_{bar}\le 2-3 \times
10^{11}M_\odot$, and a disk$+$bulge mass of $M_{disk}\approx 5\times
10^{10}M_\odot$. This allows for a considerable fraction of the total
baryonic mass to reside in a gaseous halo, partly in a hot phase at
the virial temperature $T\approx 10^6 {\rm K}$ spread out to the
virial radius $R_{\rm vir}\sim 300 {\rm kpc}$, and partly in a cooler
halo component.

Hydrodynamic simulations of galaxy formation predict that the hot MW
halo is linked to its cosmological environment and should threfore be
highly inhomogeneous. Typically, gas would be accreted into the halo
along filaments which are oriented based on the MW's position within
the cosmic web of dark matter and baryons.  Following the results
presented by \citet{bregman07b} that the MW currently retains its
separate halo, we expect filaments to be directed towards the MW's
center rather than the Local group center.  The gas in halos of
galaxies is not always heated by a virial shock near the virial
radius. Birnboim \& Dekel \citep{bd03,db06} showed analytically that
the post-shock gas is not always able to support the virial shock. If
the halo mass is below some threshold, the post-shock gas is unstable
to cooling, and the virial shock cannot propagate outwards. Rather, it
remains close to the galactic disk. The threshold mass for metal-poor
gas is $\sim 5\times 10^{11}M_\odot$ and for gas at $0.3$ of the solar
metallicity it is $\sim 10^{12}M_\odot$. At high redshifts, halos
above this mass threshold originate in rare density peaks, and such
halos typically reside in the intersections of filaments within the
cosmic web. This allows for denser filaments to avoid shock heating,
even if the smooth, diffuse halo hosts a stable virial shock. At
$z>2$, halos above $10^{12}M_\odot$ are expected to contain cold flows
within filaments embedded in a hot gaseous halo.  This picture has
been verified by different numerical simulations
\citet{keres05,ocvirk08} and by $H_\alpha$ observations
\citep{genzel06}. The MW halo is just above the above mass threshold
and it contains a hot halo. However, the filamentary structure is
still expected to exist, as suggested by \citet{keres08}, who used
cosmological {\it Smooth Particle Hydrodynamics (SPH)} simulations to
show that overdense filaments of MW-type halos are more than an order
of magnitude denser than the hot diffuse gas in which they are
embedded, although $80$--$90\%$ of the gas remains at the virial
temperature.

The hot gas in the MW halo has a cooling time that is shorter than the
halo formation time interior to some radius \citep{white91} and its
central core is expected to cool. Detailed 1D hydrodynamic simulations
find, however, that the hot halo component does not cool significantly
\citep{bdn07} for some period of time after the initial formation of
the virial shock. As the post-shock gas becomes stable, the shock
rapidly propagates from the central galaxy to the edge of the halo,
halting the gas infall and, in some cases, putting it in an outward
ballistic motion. This causes a complete shutdown of gas accretion
which lasts for a few billion years. In the absence of a heating
source, cooling would resume after this phase.  The unimportance of
``hot mode'' accretion due to cooling of hot gas and settling to the
galactic disk is also seen in 3D simulations of \citet{keres08}. Mass
limits on the galaxy indicate that even if gas had cooled, it did not
settle into the galaxy, and would be orbiting the halo as cold clumps
\citep[but see Ref.][which demonstrated that the formation of cold
clumps within the halo via a cooling fragmentation instability is
highly unlikely]{binney09}.  In that case, the cold gas in the halo
would be partially ionized \citep{maller04,kaufmann08} and yield an
even larger signature on the CMB anisotropies.

A different local contribution to the kSZ effect can arise from high
velocity clouds (HVCs). While the nature and formation scenarios of
HVCs are still under debate \citep{binney09}, we will adopt the
inference \citep{blitz99} that HVCs represent compact gas clouds with
masses of $10^6$--$10^8M_\odot$. The orbital velocities of HVCs are
typically $\gtrsim 100~{\rm km~s^{-1}}$ and they are typically found
in the vicinity of the galactic plane (at $|z|\le 10~{\rm kpc}$
(although the latter may very well be an observational selection
effect). Some of these clouds emit H$_\alpha$ radiation
\citep{putman03}, indicating the existence of ionized hydrogen.  The
gas is ionized by sources in the Galactic disk, or by hydrodynamical
heating (e.g. from the interaction between the Magellanic stream and
the hot halo gas \citep{weiner96,olano08}).  The known distribution of free
electrons in the Galactic disk yields a kSZ effect that is mostly
confined to a narrow strip of the sky \citep{hajian07,waelkens08},
where other Galactic foregrounds (resulting from synchrotron and dust
emission) dominate.

In this work we calculate the kSZ effect from the ionized gas in the
MW halo, which cover the entire sky. We proceed along two
paths. First, we model in \S \ref{sec:sources} three halo components:
{\it (i)} a diffuse hot gas at the virial temperature; {\it (ii)} cold
filaments; and {\it (iii)} high velocity clouds, and then estimate the
relative effects of these components on the CMB in \S
\ref{sec:results}. Second, we adopt an empirical approach in \S
\ref{sec:cc}, and cross-correlate various sky maps of halo gas tracers
with the CMB data from WMAP5. Finally, we summarize our results and
other possible sources of the kSZ effect in \S \ref{sec:discussion}.

\section{kSZ sources in the Milky-Way Halo}
\label{sec:sources}
Before examining the different kSZ components of the Galactic halo,
let us obtain a rough estimate for their expected magnitude.  The kSZ
shift in the CMB brightness temperature is given by \citep{sz80},
\begin{equation}
\label{eq:ksz}
\frac{\Delta T}{T}=-\int{\frac{v_R(D)}{c}\sigma_T n_e(D)dD},
\end{equation}
where $v_R$ is the radial peculiar velocity at a distance $D$ along
the line of sight (relative to the cosmic rest frame), $n_e(D)$ is the
electron number density, $c$ the speed of light and $\sigma_T$ is the
Thomson cross section.   A plausible column density $N_{e}=10^{21}~{\rm
cm^{-2}}$ of halo electrons moving at velocity $600{\rm km~s^{-1}}$
(of order the velocity of the MW relative to the CMB) yields a
brightness temperature change of ${\Delta T}/{T}= 1.2 \times 10^{-6}$,
which is detectable in current CMB measurements.

\subsection{The Milky-Way's invisible halo}
\label{sec:halo}
The baryonic mass fraction of the MW galaxy is expected to be close to
the average cosmic value, $\Omega_b/\Omega_m\approx 0.167\pm 0.008$
\citep{hinshaw09}, since the MW potential well is sufficiently deep
(and the central black hole is of relatively low mass) for feedback to
be inefficient in driving baryons out \citep{ds86}. Using highly
efficient feedback, \citet{dave09} finds that $65\%$ of the gas
remains in a MW sized halo.  By subtracting the mass of stars and gas
in the galaxy, $5$--$7\times 10^{10}M_\odot$
\citep{klypin02,shattow08}, from the total baryonic gas of $\sim
1.5$--$3\times 10^{11}M_\odot$, we infer that the halo gas has a mass
of $\gtrsim 10^{11}M_\odot$. Since the halo mass is larger than the
critical mass for the formation of a virial shock \citep{bd03,db06}, a
hydrostatic halo is expected to form. Detailed cosmological
simulations \citep{keres08} predict that a sub-dominant fraction of
the accreted baryons will be accreted in cold filaments which are
pressure confined by the hot gaseous halo.  This expectation is
supported by \citet{kaufmann08} who find that under some initial
conditions the halo gas fragments into cold and hot components, and
that most of the gas remains in the halo for $10$Gyrs.  We assume that
at least half of the halo baryons, $5 \times 10^{10}M_\odot$, reside
in the hot halo.

\begin{figure}
\begin{center}
\includegraphics[width=3.5in]{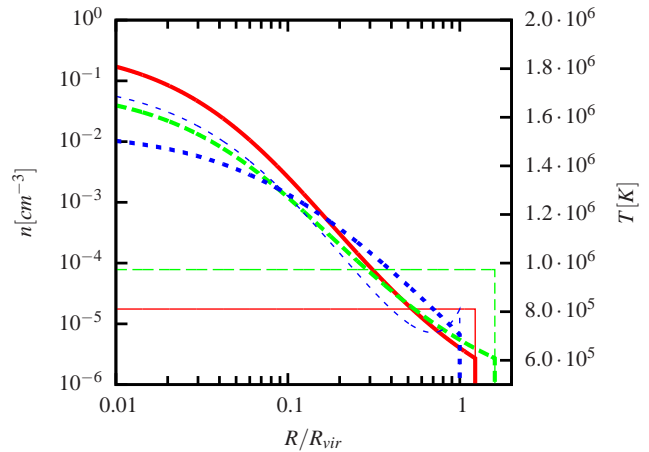}
\caption{\label{fig:profiles} Density (thick lines) and temperature
  (thin lines) of the virialized gas in the MW halo for three models:
  {\it (i)} an isothermal profile at the virial temperature $T_{\rm
  vir}$ ({\it iso}, red solid line); {\it (ii)} an isothermal
  profile at $1.2T_{\rm vir}$ ({\it 1.2iso}, long-dashed green
  line); and {\it (iii)} NFW profile with a gaseous core ({\it
  core}, dashed blue line).
The virial radius is $296{\rm kpc}$.
}
\end{center}
\end{figure}

We assume that the total mass distribution is described by an NFW
\citep{nfw97} profile with $M_{vir}=1.5\times 10^{12}M_\odot$,
corresponding to the updated circular velocity of the MW
\citep{shattow08,reid09}. Based on Refs.
\citep{klypin02,bullock01_c}, we adopt a concentration parameter of
$c=12$.  The virial radius for this mass is $296 {\rm kpc}$.  We
consider two models for the halo gas distribution in hydrostatic
equlibrium within the dark matter potential: {\it (i)} an {\bf
isothermal} (constant temperature) profile ; and {\it (ii)} a
generalized NFW density profile \citep{db08} with an inner core, in
analogy with galaxy clusters \citep{donahue06} and hydrodynamic
simulations \citep{faltenbacher07},
\begin{equation}
\rho_{gas}=\frac{\rho_s}{(1+x)^3} ,
\end{equation}
where $x\equiv r c/R_{\rm vir}$ and $\rho_s$ is a normalization
factor.  In the isothermal models, the central density of the gas is
set so that the total baryonic mass is encompassed within a radius
where the baryonic overdensity is $30$ (relative to the cosmic
average). For the {\it core} model, the gas density is normalized so
as to give the total amount of assumed mass within $R_{\rm vir}$, and
the temperature profile is fixed by the hydrostatic equilibrium
equation.  The resulting temperature profile is not far from
isothermal, with an overall temperature variation by a factor of
2. Figure \ref{fig:profiles} shows the density and temperature
profiles for the isothermal model (model ``{\it iso}'') with $T_{\rm
iso}=T_{\rm vir}=8.1\times 10^5K$ (solid red line), $T_{\rm iso}=1.2
T_{\rm vir}=9.7\times 10^5K$ (model ``{\it 1.2iso}'', long dashed
green line), and the core model (model {\it core}, dashed blue
line). Although the profiles do not differ much from each other and
they all include the same total baryonic mass within $R_{\rm vir}$, the
electron column density of a radial line-of-sight from the center
outwards 
changes significantly between $N_{21}\equiv (N_e/10^{21}~{\rm
cm^{-2}})=4.9$, $1.7$ and $0.54$ for models {\it iso, 1.2iso} and {\it
core}, respectively.

\citet{kaufmann08} performed detailed SPH simulations of MW-like
galaxies based on previous work about cooling-induced fragmentation in
galactic halos \citep{maller04}. They compared the halo cooling and
galaxy accretion properties of two MW halo models: one with the gas in
an NFW profile (their ``{\it low entropy}'' model), and the other with
a core profile (``{\it high entropy}'' model), much like our ``{\it
core}'' model, with a density of $n\approx10^{-3}~{\rm cm}^{-3}$ at a
radius of $0.1R_{\rm vir}$. They have found that while the first model
cools the gas too fast, producing a disk with $50\%$ of the total
baryonic mass, angular momentum deficiency, and a halo with a too high
soft X-ray luminosity, the second model fragments via cooling
instability into a configuration of cold clouds embedded within a hot
diffuse halo which does not suffer from these problems. A fraction
$\sim 10\%$ of the gas remains embedded in the hot halo as HVCs with a
predicted covering factor of $\approx 40\%$. The contribution of HVCs
to the kSZ signature of the MW halo will be discussed in \S
\ref{sec:hvcs}.

The rotation velocity of the hydrostatic MW halo is expected to be
small, and so we assume that the halo is moving at a constant bulk
velocity relative to the CMB \citep{kogut93}: $V_{GC-CMB}=(552.2{\rm
km~s^{-1}},l=266.5,b=29.1)$. We assume that the Sun is located at a
distance of $8.5{\rm kpc}$ from the galactic center
\citep{ghez08,reid09,shattow08}.  The off-center viewpoint creates a
quadopole, and higher spherical harmonics multipoles even for this
simple configuration.

\subsection{Filamentary infall into the Milky-Way halo}
\label{sec:filaments}
In the $\Lambda$CDM cosmology, dark matter or gas are channeled into
galactic halos through filaments which are interconnected to a cosmic
web. The gas in the filaments is cold with a temperature in the range
$\sim 10^4$--$10^5$K. As the cold gas enters the galactic halo, its
subsequent thermodynamic evolution is determined by its metalicity and
the gravitational potential well of its host halo. Previous analytic
studies \citep{bd03,db06} have found that in halos with a virial mass
below $10^{12}M_\odot$ and a metalicity of $\sim 0.3Z_\odot$, the gas
is expected to fall to the center without shocking.  For halos in a
range just above this mass threshold at redshifts $z>2$, a hot gaseous
halo is expected to co-exist together with cold, unstable
filaments. Detailed cosmological SPH simulations \citep{keres05} and
Eulerian AMR simulations \citep{ocvirk08} in 3D, have confirmed these
expectations. With the MW mass so close to the threshold mass, the
filaments are still expected to exist within the MW halo.  The
confinement of dense infalling filaments within the diffuse halo
component was investigated in detail by \citet{keres08}, who have found
that the filament gas is overdense by 1-2 orders of magnitude relative
to its hot environment and is infalling at $\sim 200~{\rm
km~s^{-1}}$. The infalling gas stops at small radii in the vicinity of
the galactic disk. Although the MW is expected to be fed by filaments,
it is not possible to predict their number, mass flux, and geometry.
In this paper, we will assume that there are either one or three
filaments, oriented radially towards the Galactic center, each having
an infall speed of $200{\rm km~s^{-1}}$, a conical opening angle of
$10^\circ$ and an overdensity of $10$ relative to the ambient hot
halo.  The direction of these filaments is chosen randomly; this
particular choice has little effect on the kSZ power spectrum (\S
\ref{sec:powers}).

\subsection{High velocity clouds}
\label{sec:hvcs}
High velocity clouds (HVCs) are observed in $21$cm HI data
\citep{blitz99} as concentrations of gas which are moving at
velocities $|V_{LSR}|>100{\rm km~s^{-1}}$ relative to the local
standard of rest (LSR) corresponding to the rotation of the Galaxy.
The origin of the HVCs is still debated in the literature.  Possible
associations include the MW disk \citep[see, e.g. Ref.][and references
therein]{wakker08}, debris from the large or small Magellanic clouds
\citep{olano08}, extragalactic origin \citep{blitz99,binney09}
\citep[but see also Ref.][which claims that the 21cm emission
originates within a distance of 100 pc]{verschuur07}. The distance and
ionization fraction of individual clouds (and therefore their size and
mass) are unknown. \citet{putman03} estimated distances of HVCs based
on their H$_\alpha$ flux and models for the emission of ionizing
radiation from the MW. They find, within the modelling and measurement
uncertainties, that most HVCs are within a distance of $\lesssim
30{\rm kpc}$. Here we use these results to calculate the inferred
electron column densities of HVCs by assuming, for simplicity, that they
are spherical homogeneous clouds. For the angular diameter of HVCs on
the sky, we adopt their typical observed diameter, $\theta=0.2{\rm
rad}$ \citep[based on WHAM and LAB
surveys; see Refs.][]{haffner03_wham,bekhti08_lab}. The total H$_\alpha$ photon
flux $f$ (in ${\rm photons~cm^{-2}~s^{-1}~sr}$) from a HVC at a
distance $D$ and angular diameter $\theta$ is obtained from the
recombination rate,
\begin{equation}
\pi \left(\frac{\theta}{2}\right)^2f=\frac{\alpha_B n_e^2 V_{HVC}}{4\pi D^2}
\label{eq:halpha}
\end{equation}
where $V_{HVC}=4\pi/3R_{HVC}^3$ is the volume of the cloud, with
$R_{HVC}=D\times\theta/2$ being its physical radius.  Here
$\alpha_B=1.17\times 10^{-13}{\rm sec^{-1}cm^3}$ is the case B
recombination rate into H$_\alpha$ photons \footnote{We use the table
for case B, low density recombination rate from \citet[\S4.2
of][]{osterbrock06} for a gas temperature of $10^4~{\rm K}$.}. The
electron column density through the center of the HVC is,
\begin{equation}
N_{e}=2R_{HVC}n_e=\sqrt{\frac{6\pi Df}{\alpha_B}}.
\end{equation}
In calculating the peculiar velocity of a HVC, we first transform the
radial LSR velocity from \citet{putman03} to the CMB frame of
reference, based on the measurements of \citet{kogut93},
\begin{equation}
V_{\rm{HVC-CMB}}=V_{\rm{HVC-LSR}}+V_{\rm{LSR-GC}}+V_{\rm{GC-CMB}} ,
\end{equation}
with $V_{\rm HVC-LSR}=v_R\hat{R}$ where $\hat{R}$ is the unit vector
in the HVC direction. The velocities have been transformed to
Cartesian Galactic coordinates in which the $x$-axis is oriented
towards the Galactic center and the $z$-axis is oriented perpendicular
and out of the galactic plane. In these coordinates,
$V_{\rm{LSR-CMB}}=(-37,-259,269){\rm km~s^{-1}}$.  

We analyse 25 HVCs from Table 1 of \citet{putman03} that include all
the parameters needed. A more complete inventory of HVCs and their
H$_\alpha$ measurements is necessary to produce foreground maps for
future CMB experiments. The full contribution of these HVCs to the sky
map will be shown in \S \ref{sec:results} but we choose, for
illustrative purpose, to consider the values from \citet{putman03} for
a single HVC, ``MS1-IIa'', which is a HVC in the Magellanic
stream. For this cloud $v_{R-LSR}=-120{\rm km~s^{-1}}$ according to HI
measurements and $-124{\rm km~s^{-1}}$ according to H$_\alpha$
measurements. Converting to the CMB reference frame, the peculiar
radial velocity of this HVC is $V_{HVC-CMB}\times \hat{R}=-376{\rm
km~s^{-1}}$.  For a distance $D=9.7{\rm kpc}$ and an H$_\alpha$ flux
of $f=407 {\rm mR}=3.24\times 10^4 {\rm photons~cm^{-2}~s^{-1}~sr}$,
we find $N_{e}=1.7\times 10^{20}{\rm cm^{-2}}$ and since the reported
value of $N_{HI}$ is $1.1\times 10^{20} {\rm cm^{-2}}$ we infer that
this cloud is partly ionized. The mass of the cloud is $1.7\times
10^6M_\odot$ and its radius is $1~{\rm kpc}$. The kSZ effect on CMB
photons passing through the center of the cloud is according to
Eq. (\ref{eq:ksz}), $\Delta T/T=1.5\times 10^{-7}$.

The HVCs of \citet{putman03} were discovered by the proximity effect
of the galactic ionizing radiation. This restricts their detections to
the innermost $30 {\rm kpc}$ around the Galactic disk. Assuming that
the origin on these HVCs is extra-galactic, there is no reason to
assume that HVCs will only exist near the galactic disk. In addition,
a comparison of the baryons in the galaxy and hot halo to the cosmic
baryonic fraction of the total halo mass indicates that there are some
missing baryons in the MW (\S \ref{sec:halo},
\citep{maller04,kaufmann08}). We therefore construct a synthetic
distribution of HVCs. We assume that they trace an isothermal density
profile, and include $2.5\times 10^{10}M_\odot$ of mass, distributed
as $250$ HVCs, with a mass of $10^8M_\odot$ each. Their direction is
drawn randomly around the MW's center, and their distance from the MW
is drawn from an isothermal density profile. The velocity has been
drawn from a Gaussian distribution with a zero mean and an {\it rms}
value, $\sigma=100{\rm km~s^{-1}}$.  A constant angular size of
$\theta=0.2 {\rm rad}$ is assumed for all HVCs. The constant angular
dimension is justified, for example, if the halo had an isothermal
profile of gas and the HVCs at a different constant temperature
(e.g. $10^4K$) were in hydrostatic equilibrium with the hot halo gas,
and the displacement of the observer from the Galactic center is
ignored. We have verified that a modest change in the radial
distribution and the angular size distribution of HVCs does not change
their cumulative kSZ effect considerably.

\section{Estimating the kSZ effect in the MW halo}
\label{sec:results}
We generate mock kSZ maps by first constructing a 3D model of the MW
halo and then numerically integrating Eq. (\ref{eq:ksz}) over
lines-of-sights in different directions. The 3D density distribution
of the halo electrons is split into a smooth spherically-symmetric
component (see \S \ref{sec:halo}), and ``objects'' which include cold
filaments (\S \ref{sec:filaments}) and HVCs (\S \ref{sec:hvcs}).  The
various versions of the {\bf halo} component are all static and
centered around the Galactic center. Their column densities (from the
Galactic center to the virial radius) change between $0.5$ to$4 \times
10^{21}cm^{-2}$ for the ``core model'' and the ``iso model'',
respectively. The {\bf filamets} are conical objects, with an opening
angle of $10^\circ$ and an overdensity of $10$ with respect to the
smooth halo. They are radially infalling at a constant speed of
$200{\rm km~s^{-1}}$. Their central column densities are 10 times
larger than that of the corresponding halo, but the actual column
density (integrated from the Sun) depends somewhat arbitrarily on
their location on the sky (see \S \ref{sec:filaments}). The {\bf
HVCs'} velocity is either taken from observations \citep{putman03} or
chosen randomly to fill the halo (typically $|v_r|~100{\rm
km~s^{-1}}$) and they have typical column densities of $\sim 2\times
10^{20}cm^2$ (\S \ref{sec:hvcs}).  The gas density and radial peculiar
velocity at specific 3D positions are then integrated along each
line-of-sight. We adopt a HEALPix \citep{gorski05}
\footnote{http://healpix.jpl.nasa.gov} pixalization scheme with
Nside$=256$ which provides sufficient resolution for the typical low
multipoles ($l<50$) that we predict (see \S \ref{sec:results}).  The
resulting kSZ sky maps are then decomposed into spherical harmonics,
and the resulting power spectrum is compared to the CMB power
spectrum, the cosmic variance noise, and the expected instrumental
noise of the {\it Planck}
satellite\footnote{http://www.rssd.esa.int/index.php?project=planck}. The
analysis is performed using the Fortran 90 HEALPix software library.

\subsection{Sky maps}
We investigate a few halo profiles as described in Table
\ref{tab:jobs}.  Models 1--3 include only smooth profiles based on the
halo models.  Models 4 and 5 show the incremental contribution of one
filament at Galactic longitude $l$ and latitude $b$ values of
$(l=259^\circ,b=30^\circ)$ (model 4), and of two additional filaments
at $(l=120^\circ,b=30^\circ)$ and $(l=20^\circ,b=-80^\circ)$ (model 5)
with respect to the galactic center. The locations of the filaments
have been chosen arbitrarily for illustrative purposes.  Model 6 adds
to model 5 the HVC component according to \citet{putman03} (\S
\ref{sec:hvcs}) that has a sub-dominant effect relative to the halo
and filament contributions. Model 7 described a synthetic HVC
population of $250$ HVCs with total mass of $2.5\times 10^{10}M_\odot$
as described in \S \ref{sec:hvcs}. Model 8 is the maximal effect model
achieved by combining a hydrostatic {\it iso} profile, $3$ filaments
and the synthetic distribution of filaments.  The maximal effect of
model 8 is $\Delta T/T \sim 8\times 10^{-6}$, which exceeds the values
of $2\mu K$ reported by \citet{hajian07,waelkens08} for the Galactic
disk.  The square of the signal-to-noise ratio is defined as,
\begin{equation}
(S/N)^2=\frac{1}{(l_{max}\!-\!l_{min}\!+\!1)}
\sum_{l_{min}}^{l_{max}}\frac{(2l+1)C_l}{N_l^{\rm
    WMAP5}},
\end{equation}
with $l_{min}=4$, $l_{max}=20$ and $N_l^{WMAP5}$ being the total error
estimate of the unbinned WMAP5 data
\footnote{taken from
  \nolinkurl{http://lambda.gsfc.nasa.gov/data/map/dr3/dcp/wmap_tt_spectrum_5yr_v3p1.txt}}.
  We show the associated $S/N$ values in the last column of Table
  \ref{tab:jobs}.  The models with synthetic HVCs in Table
  \ref{tab:jobs} have $(S/N)\sim 0.2$, making them sub-dominant
  relative to cosmic variance, although not by much. If the real clouds 
  have a higher column density of electrons than assumed here (e.g. owing to a
  non-spherical geometry), then their contribution could become
  noticeable.

 \begin{table}
 \caption{the kSZ effects of various halo models. In the HVCs column
   '-','+','*' corresponds to no HVCs, HVCs according to
   \citet{putman03}, and synthetic HVCs,
   respectively. \label{tab:jobs}}
 \begin{tabular}{|c|@{ }c@{ }|@ {}c@{ }|@{ }c@{ }|@{ }c@{ }|@{ }c@{ }|@{ }c@{ }|}
 \hline
 \hline
  \# & halo & number of & HVCs & $min(\frac{\Delta T}{T})$&
  $max(\frac{\Delta T}{T})$ & S/N \\
  \ & model & filaments & & $[10^{-6}]$  & $[10^{-6}]$& \\
 \hline
 1 & {\it iso}    & \hspace{0.5cm}0 &\hspace{0.2cm} - & $-3.0$ &  $ 3.5$&.030\\
 2 & {\it 1.2iso} & \hspace{0.5cm}0 &\hspace{0.2cm} - & $-0.9$ &  $ 1.0$&.006\\
 3 & {\it core}   & \hspace{0.5cm}0 &\hspace{0.2cm} - & $-0.5$ &  $ 0.6$&.001\\
 4 & {\it core}   & \hspace{0.5cm}1 &\hspace{0.2cm} - & $-1.0$ &  $ 0.6$&.018\\
 5 & {\it core}   & \hspace{0.5cm}3 &\hspace{0.2cm} - & $-1.0$ &  $ 1.1$&.041\\
 6 & {\it core}   & \hspace{0.5cm}3 &\hspace{0.2cm} + & $-1.0$ &  $ 1.1$&.043\\
 7 & {\it core}   & \hspace{0.5cm}3 &\hspace{0.2cm} * & $-3.0$ &  $ 5.4$&.16\\
 8 & {\it iso}    & \hspace{0.5cm}3 &\hspace{0.2cm} * & $-6.1$ &  $ 8.0$&.21\\
 \hline
 \end{tabular}
 \end{table}

\begin{figure}
\begin{center}
\includegraphics[width=3.5in]{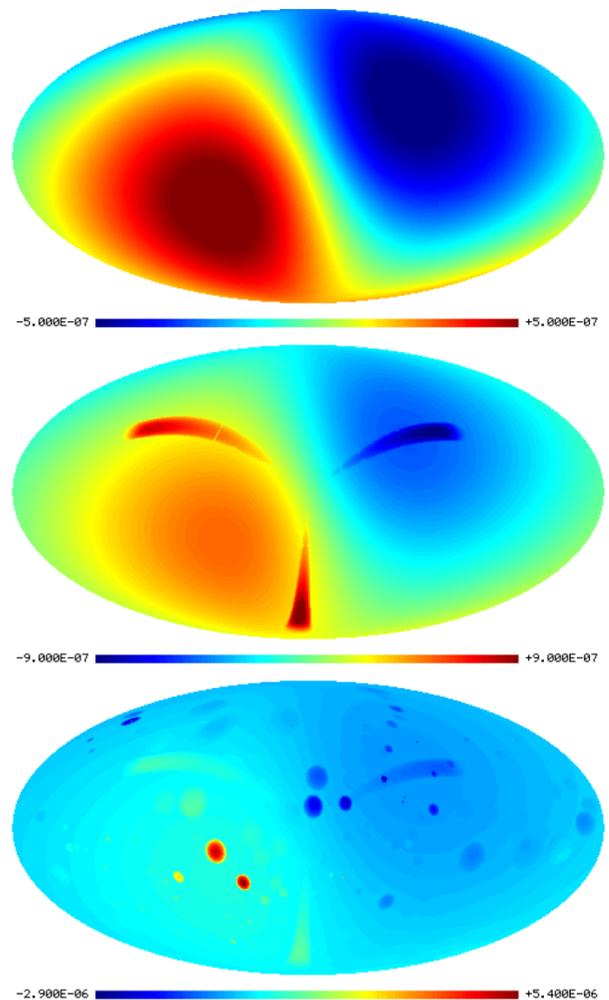}
\caption{\label{fig:skymap} Fractional changes in the CMB brightness
  temperature, ${\Delta T}/{T}$, from the kSZ effect of various models
  for the MW halo. All three halo models include a smooth component of
  the ``{\it core}'' density profile, described in \S \ref{sec:halo}
  and Fig. \ref{fig:profiles}. {\it Upper panel:} the smooth {\it core}
  model (model 3). {\it Middle panel:} {\it core} model with three
  filaments (model 5). {\it Lower panel:} {\it core} model with three
  filaments and synthetic HVCs (model 7).}
\end{center}
\end{figure}

Figure \ref{fig:skymap} shows the kSZ sky maps for a smooth halo
model, a halo model with filaments, and a halo model with filaments
and HVCs (models 3, 5 and 7, respectively) with the monopole and
dipole components owing to our motion removed. The dipole resulting
from the relative velocity of the halo electrons and the CMB is shown.

\subsection{kSZ Power spectra}
\label{sec:powers}
\begin{figure}
\begin{center}
\includegraphics[width=3.5in]{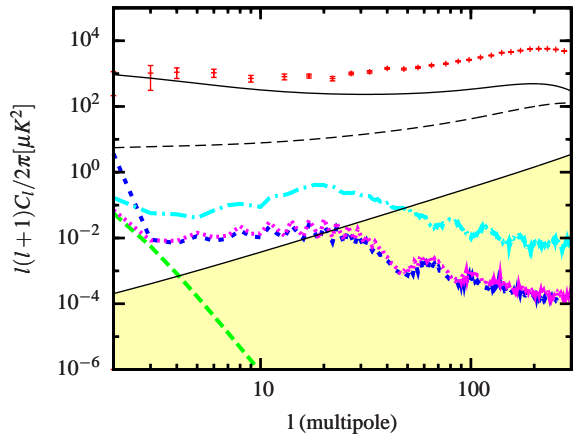}
\caption{\label{fig:powers}The CMB WMAP5 power spectrum (red curve
  with error bars) in comparison to the power spectrum of the {\it core}
  halo profile (model 3, long-dashed green line), {\it core} with 3
  filaments (model 5, dotted blue line) and {\it core} with 3 filaments
  and HVCs (model 6, purple dashed line) and {\it core} + 3 filaments +
  synthetic HVCs (model 7, cyan dot-dashed). The thin solid lines
  delineate the cosmic variance uncertainties (higher line) and
  {\it Planck}'s instrumental noise (lower). 
  The instrumental noise of the WMAP5 data is denoted by the thin, dashed line.
  Features in the shaded area
  below the instrumental noise line will not be observable with the
  {\it Planck} satellite.}
\end{center}
\end{figure}

The power spectrum is recovered from the sky maps by decomposing the
sky maps into spherical harmonics, $[\Delta
T/T](\theta,\phi)=\sum_{l,m} a_{lm}^{\rm halo} Y_{lm}(\theta,\phi)$,
and averaging over $m$ to obtain $C_l^{\rm halo}={1\over 2l+1}\sum_m
\vert a_{lm}^{\rm halo} \vert^2$ for each $l$.  Figure
\ref{fig:powers} compares the (binned) power spectra of the 5-year
WMAP data \citep{hinshaw09} to that of models 3,5,6 and 7.  For
reference, the figure also shows the statistical noise arising from
cosmic variance.  The figure also presents the expected instrumental
noise of the {\it Planck} satellite,
\begin{equation}
N^{exp}_l=\langle n_l,n_l\rangle=\Theta_{\rm{fwhm}}^2\sigma_T^2
\exp\{-l(l+1)\frac{\Theta_{\rm{fwhm}}^2}{8ln2}\} ,
\end{equation} 
with $n_l$ the noise of each $a_{lm}$, $\Theta_{\rm{fwhm}}=10^\prime$ the full width at half maximum of
the beam and $\sigma_T=5\mu K$ being the detector's sensitivity \footnote{The values
  quoted here are reasonable for the $143{\rm GHz}$ HFI instrument,
  http://www.rssd.esa.int/SA/PLANCK/docs/goal-perf-sum-feb-2004.pdf},
and the instrumental noise as reported in the WMAP5 data.

We find that the effect of the ionized MW halo is larger than {\it
Planck}'s instrumental noise for $l\le 4$. The power spectrum scales
roughly linearly with the number of filaments. The filament
contribution peaks at $l\sim 15$ and is larger than {\it Planck}'s
instrumental noise for $l\le 20$. The observed HVC's contribution to
the power spectrum is sub-dominant relative to that of the
filaments. The synthetic HVCs contribute to the power spectrum at
$l\sim 20$ and are larger than {\it Planck}'s instrumental noise at
$l\le 40$. A more complete observational census of HVCs with
H$_\alpha$ measurements (beyond the 25 HVCs known) will increase the
signal linearly in the number of HVCs and improve these constraints.
All of the kSZ power spectra lie well below the cosmic variance noise,
making them statistically insignificant for the determination of
cosmological parameters.
All of the models discussed here are below the instrumental noise of
WMAP5, and not expected to be observed.

\begin{figure}
\begin{center}
\includegraphics[width=3.4in]{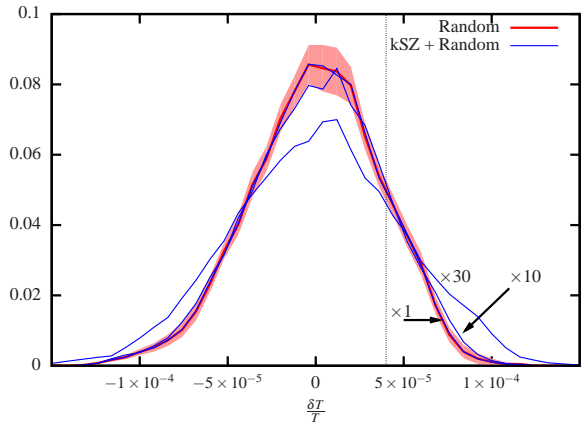}
\caption{\label{fig:gauss} The one point Gaussianity histogram of random
  maps, and random maps with model 7 (core, 3 filaments and 250 HVC)
  superimposed on it. The histogram is generated by polling the
  average value
  of ${\delta T}$ on
  disks with diameter $0.15 {\rm rad}$. The shaded area is the
  $1\sigma$ scatter of 20 such realizations. The blue lines correspond
  to the non-Gaussian contribution when model 7 is multiplied by
  $1$, $10$ and $30$. The normalization is arbitrary.}
\end{center}
\end{figure}

The kSZ contributions from filaments and HVCs are localized and should
not in general resemble a Gaussian random field. In an attempt to
quantify the non-Gaussianity of the kSZ signal we generated mock CMB
maps using the power spectrum of the WMAP5 \citep{hinshaw09} and
superimposed the results of model 7 onto these maps. We then
calculated the brightness fluctuations on different angular scales
around the HEALPix cell centers and compared a histogram of these to a
corresponding histogram of the mock CMB map without the filament and
HVC contributions. For various choices of the anglular scale and the
binning intervals, we have found that the non-Gaussian kSZ signature
of HVCs and filaments is undetectable.  If the number of the HVCs is
increased by an order of magnitude, deviations from a Gaussian
probability distribution are still virtually non-detectable. Figure
\ref{fig:gauss} compares the non-Gaussianity of model 7, when the kSZ
signal is multiplied by factors of $1$, $10$, and $30$. We quantify the
strength of the non-Gaussian signal by measuring the ``power in the
wings'' which we define as the ratio between the integral of the
histogram beyond some threshold fluctuation amplitude for the
non-Gaussian and Gaussian distributions. The cutoff chosen here is
$\frac{\delta T}=4\times 10^{-5}{\rm K}$ (thin vertical line). The
power in the wings is: $1.0025, 1.066$ and $1.439$ for the factors
of $1$, $10$, and $30$ respectively. We conclude that if the signal were an order
of magnitude larger (a factor of 10 in the column density of the
ionized gas or the velocity), then the non-Gaussianity would have been
detectable. A more rigorous three-point correlation ($f_{NL}$) test goes
beyond the scope of this paper.

\section{Cross correlations of HI and H$_\alpha$ maps with the CMB}
\label{sec:cc}
In the previous sections we utilized physically motivated models for
the halo gas including filaments and HVCs.  In this section we follow
an empirical approach and cross correlate various observational maps
that might trace the distribution of free electrons in the MW
halo. The procedure used here follows the analysis of \citet{land07}
for the Leiden/Argentine/Bonn (LAB) Galactic HI Survey
\citep{kalberla05}, with the appropriate density/velocity weighting as
discussed below. We start by describing the cross correlation of a
general sky map (that may cover only part of the sky) with the CMB in
\S \ref{sec:cross_def}. We then analyse various specific maps and
weighting schemes in \S \ref{sec:cross_res}.

\subsection{Cross-correlation formalism}
\label{sec:cross_def}
We adopt a procedure for cross-correlating a sky map with the CMB data
that resembles the one used by \citet{land07}, with the important
difference that we use a weighting scheme tailored to enhance the
kSZ signal.  Our procedure is as follows:
\begin{enumerate*}
\renewcommand{\labelenumi}{\roman{enumi}.}

\item Interpolate the sky map into a HEALPix format with
  Nside$=512$. HI maps originating from the LAB survey are in fits
  image format, and so we performed a 2D bilinear interpolation to the
  coordinates of the center of each HEALPix cell. For H$_\alpha$ WHAM
  maps \citep{haffner03_wham}, the original data is presented as
  scatter plots and we use qshed2d algorithm \citep{renka88} to
  interpolate it into HEALPix cell positions.

\item Remove the monopole and dipole of the maps by converting the
map to $(l,m)$ space and eliminating all the power in $l\le 1$, and
then converting back to real space.

\item Combine the mask of the sky map with the WMAP5 data mask KQ85
\citep{gold09} to produce the ``total mask'', and convolve the total
mask with both WMAP data (ILC and V2 map were used) and the sky map.

\item Calculate the cross-correlation of the WMAP5 map with the sky
map under consideration using the total mask.  The auto/cross
correlation of maps $a$ and $b$ is given by:
\begin{equation}
C_l^{a,b}=\frac{1}{(2l+1)}\sum_m a_{lm}^aa*_{lm}^b
\end{equation}
and the normalized cross correlation is defined as:
\begin{equation}
\hat{C_l^{a,b}}=\frac{C_l^{a,b}}{\sqrt{C_l^{a,a}C_l^{b,b}}}.
\label{eq:normcc}
\end{equation} 
A perfect correlation yields $\hat{C_l^{a,b}}=1$, a null correlation
yields $0$, and a perfect anti-correlation yields a value of $-1$ for
this parameter.

\item Since the real maps are expected to be non-Gaussian and we
  expect the nontrivial masks to introduce some correlations between
  different multipoles, we evaluate the
  statistical significance of the result by creating random mock CMB
  maps from the unmasked CMB power-spectrum, and calculating the
  masked, normalized cross-correlation $10,000$ times. The mean and
  variance of the realizations for each $l$ are then compared to the
  actual cross-correlation signal.
\end{enumerate*} 

\subsection{Observed Sky maps}
\label{sec:skymaps}

\noindent
\rmbf{LAB survey:} The kSZ weight should depend linearly on the
 electron column density and the velocity of the ionized gas
 (Eq. \ref{eq:ksz}).  We use data on the neutral hydrogen distribution
 from the 21cm LAB survey
 \citep{kalberla05,hartmann97,bajaja05,arnal00}, which provides a full
 sky map of the column density at each velocity bin relative to the
 LSR.  The data is tabulated in bins of $0.5^\circ$ in angle and
 a velocity resolution of $1.3~{\rm km~s^{-1}}$. The data covers the
 full sky, and so we use the mask of KQ85 which leaves $82\%$ of the
 sky unmasked.

The conversion of the measured HI column density $N_{\rm HI}$ to the
column density of free electrons $N_e$ requires some model
assumptions. We first consider the case where the gas is optically
thin to ionizing radiation and is distributed in a spherical shell of
a constant width $D$ around the galaxy. Ionizing radiation, either
extragalactic or galactic in origin, is impinging on the gas and
creating free electrons. In this case, a steady state balance between
the recombination and ionization rates imply,
\begin{equation}
n_e^2 D\propto n_{\rm HI}D\equiv N_{\rm HI} ,
\end{equation}
with $n_e$, $n_{\rm HI}$ and $N_{HI}$ denoting the electron density,
the HI density and the HI column density, respectively.  The LAB
survey maps $N_{\rm HI}$ as a function of velocity and so the sky
map LAB1 is constructed as
\begin{equation}
  S_{l,b}^{\rm{LAB1}}=-\int_v N_{HI}(l,b,v)^{1/2}v_{\rm{CMB}} dv ,
\label{eq:lab1}
\end{equation}
with $v_{\rm{CMB}}=v+V_{\rm{LSR-CMB}}\times \hat{r}$,
$V_{\rm{LSR-CMB}}$ from \citet{kogut93} and $\hat{r}$ the radius
vector to the point of interest $(l,b)$.

As an alternative case, we assume that the gas is optically thick to
ionizing radiation and is collisionally ionized at some constant
temperature.  In this case,
\begin{equation}
n_eD\propto n_{HI}D\equiv N_{HI} ,
\end{equation}
and 
\begin{equation}
  \rm{S}_{l,b}^{\rm{LAB2}}=-\int_v N_{HI}(l,b,v)v_{\rm{CMB}} dv .
\label{eq:lab2}
\end{equation}

\noindent
\rmbf{Outliers in HI:} In addition to cross correlating the full HI
LAB data, we also correlated the CMB with a derived map of HVCs and
outliers produced by \citet{westmeier07} by subtracting an HI halo
model \citep{kalberla07} from the data, and masking all the areas with
no extra MW component. The outlier map masks all but $12\%$ of the sky
and after convolving this mask with the CMB mask, the total mask
leaves $10\%$ of the sky unmasked. The data is expressed in terms of
the average velocity and total column density and in analogy with
Eqs. (\ref{eq:lab1}) and (\ref{eq:lab2}) we define models OL1 and OL2
as,
\begin{equation}
  \rm{S}_{l,b}^{\rm{OL1}}=-N_{HI}^{1/2}(l,b)v_{\rm{CMB}} ,
\label{eq:ol1}
\end{equation}
and
\begin{equation}
  \rm{S}_{l,b}^{\rm{OL2}}=-N_{HI}(l,b)v_{\rm{CMB}} .
\label{eq:ol2}
\end{equation}

\noindent
\boldmath \rmbf{H$_\alpha$ data:} \unboldmath The HI data, while being
extremely accurate and detailed, is an indirect tracer of free
electrons. A more direct tracer is H$_\alpha$ radiation which result
from recombinations of electrons and protons. The Wisconsin H-Alpha
Mapper \citep[WHAM,][]{haffner03_wham} provides an H$_\alpha$ map of
the sky.  Above the galactic plane, the WHAM survey has been shown to
correlate with HVCs \citep{putman03}. The WHAM survey covers $73\%$ of
the sky and the total mask amounts to $61\%$ of the sky. Assuming,
again, some typical width for the H$_\alpha$ emitting layer, $D$, and
using Eq. (\ref{eq:halpha}), the column density of ionized gas should
scale as the square root of the H$_\alpha$ flux, $f^{1/2}$, and so we
adopt the weighting scheme,
\begin{equation}
  S_{l,b}^{\rm{WHAM}}=-\int_v f(l,b,v)^{1/2}v_{\rm{CMB}} dv .
\label{eq:wham}
\end{equation}
The WHAM data is available as velocity dependent emission measure at
scattered locations, and was interpolated to the HEALPix grid
following \citet{renka88}.

\subsection{Cross correlation results} 
\label{sec:cross_res}
\begin{figure}
\begin{center}
\includegraphics[width=1.05\linewidth]{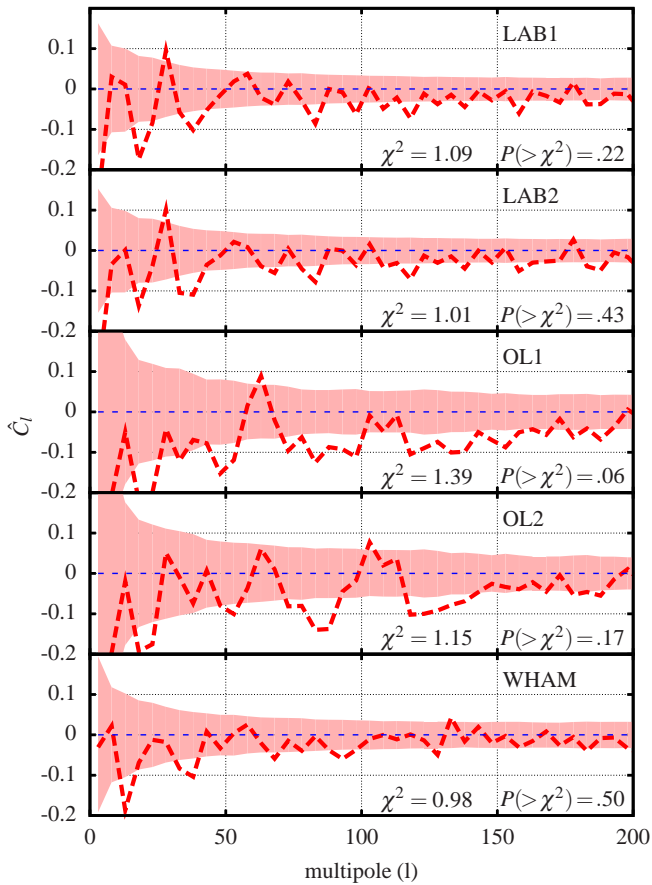}
\caption{\label{fig:cl}Normalized cross correlation as a function of
  spherical harmonic multipoles. We show the full LAB data with LAB1
  weighting (Eq. \ref{eq:lab1}, top panel), with LAB2 weighting
  (Eq. \ref{eq:lab2}, 2nd panel), HVC map with LAB1 (3rd panel), HVC
  with LAB2 (4th panel), and WHAM with weighting according to
  Eq. (\ref{eq:wham}). {\it Solid line:} cross correlation between each map
  and WMAP5 \citep{hinshaw09}. {\it Shaded area:} 1--$\sigma$ scatter levels
  from $10,000$ mock CMB realizations with the WMAP5 power
  spectrum. Data has been binned to groups of 5 consecutive multipoles
  in order to reduce the noise level. The $\chi^2$ of the cross
  correlation and the fraction of mock realizations with $\chi^2$ larger
  than the real one are marked on each panel.}
\end{center}
\end{figure}
Figure \ref{fig:cl} shows the normalized cross correlation
(Eq. \ref{eq:normcc}) of the sky maps defined in \S
\ref{sec:skymaps}. The powers here have been binned to groups of 5
$l$s in order to reduce the noise level. A null cross correlation
should have a zero correlation among different multipoles so this
averaging should not change the expected (zero) mean of the cross
correlation signal. The shaded areas on the plots mark the averaged
$1\sigma$ regions around averaged means of $10,000$ cross correlations
for each sky map with a mock CMB map of the same power spectrum.
\citet{land07} have performed a similar analysis on the LAB data, with
sky maps of column density of integrals over all or part of velocity
space (rather than our expected kSZ weightings), and have found no
significant cross correlation. We recover these results here and find
for our LAB1 and LAB2 sky maps that there is no apparent cross
correlation with the WMAP5 data. A null cross correlation is also
consistent with the OL2 and WHAM sky maps.

Sky map OL1 shows an anti correlation larger than $1\sigma$ over most
multipoles between $50<l<200$. The statistical significance of this
anti correlation was tested by defining
\begin{equation}
\chi^2=\frac{1}{(l_{max}\!-\!l_{min}\!+\!1)}\sum_{l_{min}}^{l_{max}}\left(\frac{\hat{C_l}}{\hat{\sigma_l}}\right)^2,
\end{equation}
with $l_{min}=50$, $l_{max}=200$ and $\hat{\sigma_l}$ being the standard deviation
of $10^4$ realization.
The $\chi^2$ of the actual data was then compared to the distribution of
$\chi^2$ for $10^3$ realizations. $6\%$ of the mock realizations had a
$\chi^2$ larger than the observed sky map, indicating that such an outcome is
not an extremely rare occurrence.  
Additionally, we performed various tests to eliminate
possible biases, including rotation of the OL map by an arbitrary
angle (which recovers the null result and eliminates some effects of the
mask), and checking different frequency bands of WMAP data (and
finding that the ILC and V
bands yield almost similar, statistically significant
anti-correlation). The anti correlation is probably due to
correlations of different multipoles introduced by the nontrivial mask
which covers $89\%$ of the sky, while a true kSZ
detection should have yielded a positive correlation. 
We also note that the OL sky maps were
derived from the \citet{westmeier07} maps that were produced by
subtracting an HI halo model from the LAB data. An over subtraction
could have produced an anti correlation of the type that was found. 
Finally, the LAB, WHAM and outlier maps have been searched
(non-systematically) for features that resemble some of the predicted
halo features in \S \ref{sec:sources} and no clear correlation was
found.
  
\section{Discussion and conclusions}
\label{sec:discussion}
We examined the kSZ effect from the MW halo by first constructing
theoretical models for the expected signal, and then adopting an
empirical strategy of cross-correlation observed maps of halo gas
tracers with the WMAP5 data.  Our theoretical models include smooth
hot gas, filaments of cold inflowing gas, as well as HVCs.  The
cross-correlation was analysed for the observed sky maps of HI and
H$_\alpha$ emission.

Figure \ref{fig:powers} shows that the kSZ signatures from the smooth
halo and the filaments declines sharply for multipoles above $l=4$ and
$20$, respectively, and is unlikely to be detectable considering the
orders of magnitude gap between the signal and the CMB power
spectrum. The contribution from HVCs peaks at $l\sim 20$ with an
amplitude which is $\sim 3$ orders of magnitude lower than the CMB
power-spectrum and $\sim 2$ orders of magnitude lower than the
statistical noise of WMAP5. 
However, for multipoles $l\lesssim 20$--40, the total kSZ signal from
the Galactic halo is expected to be above the instrumental noise of
the {\it Planck} satellite.  Galactic foregrounds are being routinely
masked or cleaned in the analysis procedure, often based on their
unique dependence on photon frequency or Galactic latitude.  However,
the kSZ effect of the MW halo covers the entire sky and has no
frequency dependence relative to the blackbody spectrum, and so its
removal cannot be accomplished in the standard linear combination
(ILC) maps. The halo effect is larger by an order of magnitude than
the kSZ signature of the MW disk \citep{hajian07,waelkens08}, and is
not confined to a narrow strip around the galactic plane -- which is
often masked out in a cosmological analysis of the CMB
anisotropies. Our theoretical prediction for the HVC contribution to
the kSZ effect is based on H$_\alpha$ observations of these HVCs
\citep{putman03} together with a number of model assumptions. In
particular, we assume the HVCs to be spherical objects with constant
angular size, optically thin, and in ionization/recombination
equilibrium with the galactic radiation field. Better understanding of
the nature, location and geometry of these HVCs is required for a more
robust assessment of their kSZ effect. The modest sample of 25 HVCs
with known distances \citep{putman03} is incomplete, and since the kSZ
signal depends linearly on the number of HVCs, a more complete data
base is necessary. We estimated the effects of 250 such HVCs by
randomly picking them at larger radii. The HVCs are expected to leave
a distinct non-Gaussian signature on the CMB sky. Although our
preliminary tests have found this signature to be statistically
insiginificant, the future discovery of more HVCs with known distances
might add a non-negligible non-Gaussian contamination to the
primordial CMB signal.

The kSZ and thermal SZ effects caused by large scale structures in our
local environment has been tested by \citet{dolag05} using constrained
realization simulations. They have found a kSZ signal comparable to
the one presented in this paper, but at larger multipoles. The thermal
component they predict is even larger.  The local group was proposed
as a possible source for the thermal Sunyaev-Zel'dovich (tSZ) effect
\citep{suto96}. The early proposals for the dipole and quadropole tSZ
amplitudes of the local group are unlikely to be real based on
measurements of halos of other galaxy groups \citep{pildis96}. The kSZ
signature of the local group would lead to a compact spot in the
direction of the Andromeda galaxy (i.e. the direction of the center of
mass of the local group), with a diameter of $\sim
20^\circ$. Unfortunately, the detection of this spot is highly
challenging.

Our empirically-calibrated sky maps of the kSZ signal are based on the
HI data \citep{kalberla05}, the HI HVCs and outliers distribution
\citep{westmeier07}, and the H$_\alpha$ emission maps
\citep{haffner03_wham}. We examined two algorithms for converting HI
maps to electron column densities: {\it (i)} gas which is
optically-thin to ionizing radiation, and {\it (ii)} gas with a
constant ionization fraction. We have found no statistically
significant cross-correlation between the HI and H$_\alpha$ data and
the CMB data.

We conclude that the kSZ signal from HVCs imprints pattens on the CMB
sky which are detectable but are statistically insignificant in the
cosmological context.  The kSZ effect could potentially be used to
study the content of the Galactic halo. A detection of non-Gaussianity
with the spectral features predicted here ($l\leq 20$) would
correspond to objects that are either very big or within the local
group. Fine tuning of the kSZ effect
could be invoked to explain some non-Gaussian features that are seen
in the WMAP data. The cold spot \citep{vielva04} can be explained by a
dense HVC cloud receding from us at a relatively high speed.  The
details of such a cloud and the statistical probability that such a
cloud exists is left for further investigation.

While cross correlation between various observable sky maps and CMB
data is routinely done, we derive here a prescription to estimate the
specific contribution from the kSZ effect, that incorporates the
column density and velocity in a self consistent way. If a sky map
traces the distribution of ionized gas, but its line of sight
velocities are not highly correlated, a cross correlation between that
map and the CMB will fail to detect the kSZ feature unless the
velocities are takes into account in a manner described in \S
\ref{sec:cc}. Reversely, a situation in which a sky map convolved
according to \S \ref{sec:cc} prescriptions that produces a correlation
while the naive luminosity or column density map fails to yield one
would indicate that kSZ is observed at high probability.  Finally, a
more complete future compilation of HVC data could help in creating
masks that remove this foreground contamination as well as constrain
the unknown physical properties of HVCs.

\begin{acknowledgments}
We thank Tobias Westmeier for making the HVC and outlier map available
to us digitally, and Matt McQuinn, Anthony Stark and Matias
Zaldarriaga for useful discussions and comments.  The Wisconsin
H-Alpha Mapper is funded by the National Science Foundation.  Some of
the results in this paper have been derived using the HEALPix
\citep{gorski05} package.  This work was supported in part by the
Harvard University.

\end{acknowledgments}

\bibliography{yuval}

\end{document}